\documentclass{appolb}
\usepackage{graphicx}
\usepackage[backend=biber, sorting=none]{biblatex}
\bibliography{Mooney_STAR_overview}

\usepackage{xspace}
\newcommand{\snn}{\ensuremath{\sqrt{s_{\mathrm{NN}}}}\xspace}
\newcommand{\pp}{\ensuremath{p}+\ensuremath{p}\xspace}

\begin{document}
\title{STAR Experimental Overview%
\thanks{Presented at the XXXII Cracow Epiphany Conference}%
}
\author{Isaac Mooney, for the STAR Collaboration \\
isaac.mooney@yale.edu 
\address{Wright Laboratory, Yale University, New Haven, CT; \\ Center for Frontiers in Nuclear Science, Stony Brook University, NY}
\\[3mm]
}
\maketitle
\begin{abstract}
We highlight some of the STAR collaboration's results on heavy-ion collisions from the past year, addressing many open questions related to the strong interaction under extreme conditions. Topics presented include jet and quarkonium modification in quark-gluon plasma (QGP); collective dynamics of QGP constituents; low-energy and small-sized collisions; and vector meson production from ultraperipheral, photonic ion collisions. Lastly, we end with a conclusion and outlook toward the data-analysis era. 
\end{abstract}
  
\section{Introduction}
The standard picture of heavy-ion collisions is well-developed, although there remain important open questions to be addressed. At Epiphany 2026, STAR presented results which attempted to answer some of these questions related to the scientific program of the conference. Jets and heavy-flavor particles were studied in order to assess the microscopic nature of the QGP, and its response to the passage of high-energy probes. Measurements of collective dynamics of produced particles gave insight into bulk properties of the medium, such as viscosity and temperature. Results from low-energy collisions were assessed to determine whether there exist critical fluctuations indicative of a critical point in the QCD phase diagram at high baryon chemical potential, $\mu_{B}$. The search for the smallest system in which the QGP is formed was advanced by recent results from $p$+Au and O+O collisions. And finally, heavy-ions were used instead as clean electromagnetic probes in ultraperipheral collisions in order to improve our understanding of the distribution of matter and energy in nuclei.\par

\section{The detector} \label{sec:dec}
The Relativistic Heavy-Ion Collider (RHIC) has provided STAR with more than ten collision systems, from proton-proton (\pp) to uranium-uranium (U+U) with more than fifteen energy configurations, from \snn$ = 3$ GeV (fixed target) Au+Au collisions to $510$ GeV \pp collisions. In order to take advantage of this wide range of collisions, STAR is a general purpose detector with numerous subsystems offering a wide range of measurement capabilities, and has undergone more than twenty upgrades in its more than twenty-five year lifespan. The main midrapidity subsystems include: a Time Projection Chamber (TPC) for charged-particle momentum, particle identification, and collision centrality determination with extended rapidity coverage to $|\eta| < 1.5$ after an inner TPC upgrade in 2019; the Barrel Electromagnetic Calorimeter (BEMC) which measures energy deposits of electromagnetically interacting particles and acts as a fast online trigger; and a Time of Flight (TOF) detector which measures the velocity of particles originating from the collision to determine their mass and subsequently their PID, and mitigate pileup. At large rapidity, the Vertex Position Detector (VPD) and Zero Degree Calorimeter (ZDC) act as minimum bias triggers, while the former also provides precise vertex reconstruction and the latter, luminosity monitoring. Many upgrades and subsystems are not mentioned here, including the addition of forward tracking, electromagnetic and hadronic calorimetry in 2022. \par

\section{Microscopic nature of the QGP} \label{sec:HP}
As quarkonium, a $Q\bar{Q}$ bound state, propagates through the QGP, it can dissociate and recombine due to dynamical interactions with the medium \cite{Hitschfeld, QuarkoniumTheory}
 as well as due to static screening of the $Q\bar{Q}$ potential \cite{MatsuiSatz}. The expected result is one of increased suppression for excited states with lower binding energies such as the $\psi (2S)$ and $\Upsilon (2S)$. Given that the relative balance of dissociation and regeneration depends strongly on collision energy, a measurement of the relative yields of quarkonium states in an energy regime between that of SPS and LHC experiments by STAR helps constrain the theoretical description of quarkonium-medium interactions including transport coefficients. The results \cite{STARsequentialsuppression} from Zr+Zr and Ru+Ru (isobar) collisions at \snn$ = 200$ GeV amount to the first observation of sequential suppression of charmonium at RHIC.
Comparing the double ratio of $\psi (2S)$ and $J/\psi$ yields between A+A and \pp collisions, we see a hint of increased relative suppression of excited states with increasing number of participants, $\left \langle N_{\mathrm{part}} \right \rangle$. The data from STAR are consistent with those from the SPS \cite{NA50sequentialsuppression} and LHC \cite{ALICEsequentialsuppression}
and with models with different implementations of charm quark and quarkonium evolution.
Moreover, comparing to results from $p$+Au collisions,
 where cold nuclear matter (CNM) effects are dominant, we observe additional suppression, consistent with the presence of the QGP. \par
Although quarkonium measurements give insight into the medium's effect on hard probes, the effect of hard probes on the medium is less understood. A number of proposed mechanisms may lead to a so-called medium response, including a wake effect which may push medium particles to coalesce at the peripheries of jet cones \cite{AMPTstudy}, and thermal-shower recombination \cite{HwaYangThermalRecomb} by which quarks from the parton shower may hadronize with those from the medium. In the former case, a possible enhancement of the baryon-to-meson ratio as a function of distance from the jet axis, $\Delta r$, was proposed as a signature of the effect. STAR has recently measured this ratio in Au+Au collisions at \snn$ = 200$ GeV \cite{GabeThesis}, using its particle identification capabilities and a novel approach to subtract both uncorrelated and correlated background. No monotonic increase in the ratio is observed within uncertainties, despite a hint of enhancement in one $\Delta r$ range. \par 
Another approach to the search for medium response in jets uses colorless (e.g.\ $\gamma$ and $Z$) bosons as minimally interacting particles which set a reference for the away-side recoiling colored parton (after e.g.\ gluon Compton scattering) which interacts with the QGP. STAR recently measured, in Au+Au collisions at \snn$ = 200$ GeV, the acoplanarity of jets recoiling from high-momentum photons and neutral pions \cite{acoplanarity} -- that is, the degree to which the jet axis differs from the leading-order expectation of $\Delta \phi = \pi$ compared to the trigger particle. Comparing between $\gamma$+jet and $\pi^{0}$+jet samples allows for a study of the effect of variation in initiator color and medium path-length. Results for jet radius $R = 0.2$ and $R = 0.5$ are presented, which allows for a distinction to be made between Molière scattering of the hard parton with quasi-particles in the medium and collective medium response effects. A significant radius dependence is observed, disfavoring the Molière scattering hypothesis. The data, differential in jet $p_{\mathrm{T}}$, $R$, and $\Delta \phi$, are not qualitatively described \emph{in toto} by any model, which offers an opportunity for further theoretical development.
\section{Bulk properties of the QGP} \label{sec:bulk}
Previously, measurements of flow coefficients with respect to an $n$th-order symmetry plane called the flow plane have been used to understand the collective response of the QGP to initial geometric conditions. However, it has been proposed \cite{DecorrelationsFewAuthor} that the flow planes may be sensitive to effects such as longitudinal energy density fluctuations, hydrodynamic fluctuations, and others, which would result in a flow plane decorrelation as a function of rapidity. In the same reference, a four-plane cumulant,
\begin{equation}
T_{n}\{ba;dc\} = \left\langle\left\langle\sin \left[n\left(\Psi_{n}^{b} - \Psi_{n}^{a}\right)\right]\sin \left[n\left(\Psi_{n}^{d} - \Psi_{n}^{c}\right)\right]\right\rangle\right\rangle
\end{equation}
was proposed to naturally suppress non-flow contributions, using four $n$th-order flow symmetry planes, $\Psi_{n}^{i}$, calculated on subevents in different subdetectors, labeled $a$ - $d$ above. Unlike the factorization ratio observable $r_{n}$, $T_{n}$ is also insensitive to the effect of flow \emph{magnitude} and random-walk-like decorrelations. It appears from STAR's measurement \cite{DecorrelationsSTAR} of $T_{2}$ and the relatively small values it attains, that the so-called torque effect is negligible, and instead that random-walk-like decorrelations are a dominant contributor to $r_{2}$ in isobar collisions at \snn$ = 200$ GeV, giving insight into the three-dimensional structure of the QGP. \par
In addition to the anisotropic flow coefficients, another observable which has informed our understanding of the response of the medium to initial internal pressure gradients has been the radial flow, $v_{0}$. However, unlike the anisotropic flow, $v_{n > 0}$, event-by-event fluctuations in the radial flow have not previously been studied. Recently, an observable, 
\begin{equation}
v_{0}(p_{\mathrm{T}}) = v_{0,\mathrm{int}}^{-1} \frac{\left\langle \delta n \left(p_{\mathrm{T}}\right) \delta \left[p_{\mathrm{T}}\right]\right\rangle }{\left\langle n \left(p_{\mathrm{T}}\right) \right\rangle \left\langle \left[p_{\mathrm{T}}\right] \right\rangle}
\end{equation}
representing a normalized covariance between the event-averaged mean transverse momentum and the spectrum shape at a given transverse momentum, was proposed \cite{RadialFlowFluctuations}.
This observable now allows for study of the effect of fluctuations in the collision overlap size and the subsequent hydrodynamic response. When additionally scaled by the integrated $v_{0}$, the influence of initial conditions is reduced, allowing for isolation of the collective behavior of the medium to a given initial profile. STAR reports a measurement of $v_{0} (p_{\mathrm{T}})$ in Au+Au collisions at \snn$ = 200$ GeV, and when scaling by $v_{0}$ observes the collapse of all centrality ranges onto a common curve, indicating similar hydrodynamic response, as has been previously observed with the anisotropic flow coefficients. Comparison to a hydrodynamic model with both shear and bulk contributions shows that as expected the data are sensitive to the bulk viscosity. \par
In addition to the response of the QGP to internal conditions, STAR has also made measurements attempting to understand the response of the QGP to external conditions, such as the immensely strong ($\sim 10^{18} \mathrm{G}$) magnetic field generated by charged spectators in heavy-ion collisions. If an imbalance of chirality exists in the initial state, the magnetic field can induce an electric current in the medium, in the hypothesized locally-parity-violating chiral magnetic effect (CME). By suppressing flow-related backgrounds with a novel event-shape selection technique and projection to zero estimated $v_{2}$, STAR's measurement of the multi-particle correlator $\gamma^{112} = \left\langle \cos\left(\phi_{1} + \phi_{2} - 2\psi_{\mathrm{RP}} \right) \right\rangle$ and its difference between opposite- and same-sign particles is robust to these backgrounds.
The observable is measured \cite{CME1, CME2} for multiple Au+Au collision energies, and in mid-central collisions (20-50\%), STAR observes a finite $\Delta\gamma^{112}$ with greater than $5\sigma$ significance for four energies (between 10 and 20 GeV) combined, which may indicate some impact of the initial magnetic field on QGP evolution, although more theoretical input is needed before final conclusions can be drawn. The signal is consistent with zero at 200 GeV, where the magnetic field may exist too briefly to make a sustained impact. \par
The QGP has also been measured to respond to the immense vorticity ($\sim 10^{3} \hbar$) present in non-central collisions, using e.g.\ global polarization of hyperons such as $\Lambda$ \cite{HyperonGlobalPolarization}. Recent results from STAR \cite{STARHyperonIsobar} examine this global polarization in a new, smaller system, isobar collisions, comparing to the previous results in Au+Au collisions. The new results are consistent between systems within uncertainty, despite a prediction of larger polarization in smaller systems by models due to the shorter lifetime and correspondingly less system evolution away from the highly vortical initial configuration. The data are also consistent with a hydrodynamic model assuming polarization due entirely to $s$-quarks, pre-hadronization. In the same study, STAR also attempted to address whether magnetic field spin coupling induced a splitting between the results for $\Lambda$ and $\bar{\Lambda}$, but the effect was demonstrated to be negligible with current precision. \par
Finally, STAR has recently demonstrated \cite{STARshapes}
that using our modern understanding of hydrodynamic behavior of the QGP as an input, it is possible to derive from the final state flow observables the initial conditions, including nuclear shape. By comparing nuclei with different initial shapes, for example prolate gold, and oblate uranium, it was possible to extract nuclear shape information, such as the quadrupole deformation and triaxiality parameters. New measurements \cite{STARshapes2}
including higher-order flow observables such as $\left\langle v_{3}^{2}\right\rangle$, extend the previous work and provide the first experimental evidence of octupole deformation of uranium from high-energy collisions.\par

\section{Search for a critical point in the QCD phase diagram} \label{sec:CP}
STAR's low-energy data are reported elsewhere in these proceedings \cite{Zbroszczyk}. 
\section{Search for QGP formation in small collision systems} \label{sec:SS}
Although it has been definitively demonstrated that QGP is formed in collisions of large ions such as gold, the smallest collision system which can form a droplet of QGP -- that is, thermally equilibrated, deconfined partonic matter -- remains to be determined. It was previously assumed that \pp and $p$+A collisions were incapable of forming QGP, hence their use as vacuum and cold-nuclear-matter baselines for comparisons to heavy-ion collisions where QGP was expected to form. However, recently, many of the hallmarks of QGP formation have been measured in high-multiplicity \pp and $p$+A collisions, with the notable exception of jet quenching, hypothesized to possibly be due to the proposed medium's minimal spatiotemporal extent. Another recent development was the collision of oxygen nuclei at RHIC in 2021.
 Comparing O+O and d+Au collisions, the latter has a higher initial-state eccentricity, while the two have similar triangularities. This difference is observed \cite{v2v3smallsystems} to carry to the final state flow parameters resulting in smaller $v_{2}$ for O+O collisions and similar non-zero $v_{3}$ between the two systems, with both systems well-described by hydrodynamic models, suggesting a collective response to initial geometry. \par
Another traditional indication of QGP formation is strangeness and baryon enhancement. STAR has measured the ratio of $\Omega$ ($sss$) baryons to $\phi$ (predominantly $s\bar{s}$) mesons in O+O collisions. Because thermal production is the dominant strangeness production mechanism in the presence of QGP, the enhancement of yields compared to \pp collisions is an indication of QGP formation. A similar argument holds for baryon production due to coalescence and its enhancement. Examining multi-strange hadrons increases QGP production dominance, increasing the signal. STAR observes similar enhancement to that in isobar collisions when compared at similar $N_{\mathrm{part}}$, indicating a possible thermal contribution in this system as well. \par
Lastly, as mentioned above, there have been no previous indications of jet quenching in small systems before the O+O data were taken. STAR reports a measurement in this system \cite{Sijie} of trigger-associated hadrons and jets which dramatically reduce uncertainty compared to inclusive measurements by removing dependence on the modeling of the number of binary collisions, $N_{\mathrm{coll}}$, in normalizing per trigger. The ratio between central and peripheral collisions of the associated hadrons and jets shows a suppression in the most central collisions compared to both unity and to a theoretical baseline with nPDF effects but no quenching implementation. The recoil hadron suppression, for example, in the most central collisions differs from unity with more than $5\sigma$ significance, while the near-side is consistent with unity, which is expected if there is a bias to production of the high-momentum trigger close to the surface of the hypothetical fireball. This result is suggestive of jet quenching in O+O collisions, especially when taken together with the other evidence reported by STAR in O+O collisions. \par
 
 \section{Probing matter and energy distributions in cold nuclei} \label{sec:UPC}

Ultraperipheral collisions offer a clean probe of the initial state of nuclei, taking advantage of the Weizsäcker-Williams photons emitted by passing nuclei \cite{EPA-VMD} to map the gluon distribution in the nucleus, for example. One approach is the use of coherent vector meson production: $\gamma + A \rightarrow J/\psi + A$. As the nucleus is probed coherently, it remains intact, and the resulting cross section measurement of the $J/\psi$ is related to the gluon density $g(x,Q^{2})^{2}$ of the nucleus as a whole, and any modification due to shadowing or saturation effects present in heavy ions. Comparing isobar and Au+Au collisions allows for an investigation of nuclear size dependence which ultimately varies the gluon density and saturation scale, as each of these is proportional to $A$. The result from STAR indicates suppression of the charge-scaled cross section with respect to a free-nucleon baseline, and a hint of increased production in Au+Au compared to Zr+Zr and Ru+Ru collisions. \par
Spin interference is an observable which allows for a more differential look beyond the overall gluon density to the gluon \emph{distribution} within the nucleus. In exclusive vector meson production, which can include incoherent production involving a breakup of the nucleus contributing to the results at higher transverse momentum, the vector meson inherits linear polarization from the quasi-real photon. This may be inherited by the daughters in the form of a modulation of the angle between the parent and daughters with strength $|a_{2}|$, although it is only due to quantum interference that the effect survives when averaged over an entire sample of events. STAR presents the first measurement of exclusive photoproduction of $J/\psi$ in UPCs. The negative, non-zero modulation demonstrated by the data are in agreement with a color-glass condensate calculation at low-$p_{\mathrm{T}}$, but the $p_{\mathrm{T}}$ dependence is not captured by the model for both isobar and Au+Au collisions, indicating that further theoretical developments are necessary. \par

\section{Conclusion and outlook} \label{sec:end}

In addition to the interesting results presented above, STAR has recently taken its final Au+Au collisions at top energy, in runs 23 and 25. These runs yielded 9.4 billion minimum-bias and 
$24 \mathrm{\ nb}^{-1}$ of high-luminosity and high-$p_{\mathrm{T}}$-triggered events, recorded with STAR's recent upgrades e.g.\ to the TPC for improved precision tracking, the forward upgrade to large-rapidity capabilities including tracking and electromagnetic and hadronic calorimetry, and more. This will result in improved acceptance, reduced uncertainties, and increased kinematic reach and overlap with the LHC. Additionally, STAR recorded Au+Au fixed-target data for 4, 4, and 1.5 days at \snn$ = 4.5$, $4.2$, and $5.2$ GeV respectively to complete the Beam-Energy Scan II mission, and recorded an additional five days of O+O data, ending February 6, 2026.
All data on tape will be analyzed and published over the course of roughly another decade, meaning that STAR's contribution to our understanding of the questions addressed at Epiphany 2026 is far from complete. \par

\printbibliography

\end{document}